# Characterizing interdisciplinarity in drug research: a translational science perspective

Xin Li (0000-0002-8169-6059)

School of Medicine and Health Management, Tongji Medical College, Huazhong University of Science and Technology, Wuhan 430030, Hubei, China

School of Information, The University of Texas, Austin 78701, Texas, U.S.A.

Xuli Tang* (0000-0002-1656-3014)

School of Information Management, Central China Normal University, Wuhan 430079, Hubei, China

School of Information, The University of Texas, Austin 78701, Texas, U.S.A.

* Corresponding concerning this article should be addressed to Xuli Tang (xulitang@whu.edu.cn). The address of corresponding author is School of Information Management, Central China Normal University, Wuhan 430079, Hubei, China.

## Abstract

**Background:** Despite the significant advances in life science, it still takes decades to translate a basic drug discovery into a cure for human disease. To accelerate the process from "bench-to-bedside", interdisciplinary research (especially research involving both basic research and clinical research) has been strongly recommend by many previous studies. However, the patterns and the roles of the interdisciplinary characteristics in drug research have not been deeply examined in extant studies.

**Objective:** The purpose of this study was to characterize interdisciplinarity in drug research from the perspective of translational science, and to examine the role of different kinds of interdisciplinary characteristics in translational research for drugs.

**Methods:** In this paper, we analyze the interdisciplinary of drug research from a novel bibliometric perspective, i.e., a translational science perspective. 18 FDA approved drugs belonging to 6 categories are used as the research proxies. We propose four bibliometric indicators (i.e., Diversity of Research, Symmetry of Research, Persistence of Research and Stability of Research) for characterizing interdisciplinarity in drug research at both the macro and micro levels. We also design a bibliometric indicator (i.e., Translation Intensity) for quantifying the result of a drug's translation. Correlation analysis was also used to examine the relationships between these interdisciplinary characteristics of drug research and the translation intensity of drugs.

**Results:** Multidisciplinary research (especially the research involving both basic and clinical disciplines) is still rare for all the 18 drugs and they generally occurred later than intra-disciplinary research in the timelines of drug research. The four interdisciplinary characteristics of drug research and the translation intensity of drugs both vary according to drugs. At the macro level, the diversity and symmetry of research have evident positive relationships with the translation intensity of drugs. At the micro level, the persistence of all seven kinds of research positively correlated with the translation intensity of drugs; and the persistence of multidisciplinary research have stronger relationships with the translation intensity of drugs than those of intra-disciplinary research except for the research within the cell/molecular discipline (CC research), which showed the most impressive correlations with the translation intensity of drugs. In addition, the stability of all kinds of research did not show a significant influence on the translation intensity of drugs.

**Conclusions:** Interdisciplinary research engaging both basic science and clinical science should be encouraged in translational research for drugs. The basic research within the cell/molecular discipline needs more persistence to shorten the translational lags in drug research. Moreover, the methodology in this paper showcases a feasible way to characterize interdisciplinarity of research and measure the results of translation in drug research, and it can be adopted and improved for other domains, such as vaccine, medical devices.

# 1. Introduction

## 1.1. Background

In spite of ongoing advances in science and technology, it has taken significantly longer than expected for a basic drug discovery to be translated into a treatment for human diseases. Only around 0.1% of the new drug candidates can finally receive approval from the U.S. Food and Drug Administration (FDA), with a cost of $2.6 billion for each approved drug. Almost 90% of the drug candidates failed before they ever tested in human trials, which is known as "the valley of death"; and around 50% of the drugs that have entered into clinical trials died in their phase III trials (Li et al., 2020; Seyhan, 2019). The failure rate of clinical trials of drugs has actually increased over recent years despite the increasing predictability of tests on cell/molecular or animal models (Leenaars et al., 2019). The main reasons are inadequate effectiveness or badly side effects (Waring et al., 2015). Therefore, it is crucial to improve the success rate of the translation of drugs and cut down the translation lags (Ogier et al., 2019; Parrish et al., 2019).

Most of the previous studies on translational research for drugs have focused on the roadblocks hindering the success of "bench-to-bedside" translation, such as insufficient research funding (Hörig et al., 2005), imperfect reward system (Fishburn, 2013), knowledge gap between basic and clinical scientists (Rocca, 2017), misuse of statistical methods (Vidgen & Yasseri, 2016), difficulty in cohort recruitment (Segura-Bedmar & Raez, 2019), irreproducibility of basic experiments (Jarvis & Williams, 2016), and insufficient understanding of translational science (Seyhan, 2019). To clear these roadblocks, interdisciplinary research, especially those involving both basic and clinical science, has been highly recommended as one of the solutions to accelerate drug translation (Ameredes et al., 2015; Bahney et al., 2016; Rocca, 2017). On the one hand, drug translation is an intricate task that requires a range of diverse skills, such as pharmacology, epidemiology, statistics, genetics, clinical studies and computer science (Kumar & Sattigeri, 2018). Interdisciplinary research has shown the effectiveness to breed and amplify innovations during the process of translation (Xu et al., 2015) and improve the quality and reproducibility of research (Barba, 2016), by allowing scientists in various disciplines to exchange innovative ideas and share their resources and experience (C. Zhang et al., 2018). Moreover, interdisciplinary collaborations between researchers from various organizations including funding agencies and government departments bring more funding and higher clinical impact to translational research (Gil-Garcia et al., 2019). But on the other hand, other studies have indicated several drawbacks of interdisciplinary research and collaborations for translational medicine, such as communication gaps (Grippa et al., 2018), time-consuming (Bu, Ding, Xu, et al., 2018), leadership (Folkman et al., 2019), and intellectual barriers (Banner et al., 2019). Establishing a persistent and stable interdisciplinary research team is definitely a challenge (Seyhan, 2019).

Hence, interdisciplinary research arguably plays many roles in the translation of drug discoveries, yet it has been a subject of few quantitative studies in the translational research for drugs. Besides, previous studies on interdisciplinary research focused mainly on the perspectives of journals (L. Zhang et al., 2016), articles (Leydesdorff & Ivanova, 2021), and authors (Bu, Ding, Xu, et al., 2018). However, in the process of translating drug discoveries into therapies, there is a variety of research types. In particular, the biomedical studies can be classified into seven different categories by using the translational triangle of biomedicine originally proposed by (Weber, 2013): animal related research (A), cell/molecular related research (C) and human related research (H) and the combinations of these three (i.e., AC, AH, CH, ACH). In this paper, we divided drug research into the corresponding seven categories: (1) A-A; (2) C-C; (3) H-H; (4) A-C; (5) A-H; (6) C-H; and (7) A-C-H. (1), (2) and (4) are research within basic science; (3) research within clinical science; (5), (6) and (7) are research involving both basic science and clinical science. To our best knowledge, the patterns and the roles of the different kinds of research in translational process for drugs, which can provide us insights on how to shorten the translation lags in drug research and development, have not been deeply examined in the extant studies. For example, would more persistent and stable interdisciplinary research involving both basic science and clinical science lead to better results in the translation of drugs? The more diverse the research, the more likely the translation of drugs will be successful?

**1.2. Objective**

To analyze these patterns and roles, in this paper, we first characterized the patterns of interdisciplinarity in translational research for drugs at both the macro and micro levels. Specifically, for the macro level, we employed the entropy concept to quantify the diversity of drug research by treating the relationships between research categories and the related articles of a drug as a bipartite graph (Corrêa Jr. et al., 2017; Lu et al., 2019). Then, the normalization of the diversity of research was used to calculate how the evenness of the distribution of different categories of drug research is, that is, the symmetry of research. For the micro level, we further extended two bibliometric indicators originally proposed for measuring the persistence and stability of the collaboration between an author pair by (Bu, Ding, Liang, et al., 2018; Bu, Murray, Ding, et al., 2018), to quantify the persistence and stability of each category of drug research.

To examine the role of different kinds of research in translational process of drugs, in this paper, we proposed a bibliometric indicator (translation intensity) to quantify the result of a drug's translation, which is based on the approximate potential to translate of a biomedical article proposed by (Hutchins et al., 2019). Then, correlation analyses (including Pearson and Spearman correlation analysis) were used to explore how the aforementioned interdisciplinary characteristics of different drug research had influenced the translational intensity of drugs.

We selected 18 approved drugs from the FDA website (http://fda.gov/Drugs) as the research proxies. These drugs belong to 6 different drug classes, which are common and well-studied (Kissin & Bradley, 2012). Each drug class includes one first-in-class drug (FICD) and two follow-on drugs (FOD) since it is significant to identify the differences in the patterns between real drug breakthroughs and the follow-on drugs (Kissin & Bradley, 2012).

## 2. Related work
### 2.1 Studies on characterizing the interdisciplinarity of scientific research

Interdisciplinary research, which integrates ideas, methods, techniques, theories, or knowledge from multiple disciplines (Rosenfield, 1992), has been considered to have a high probability of success. Understanding how interdisciplinary research affects scientific success can be enhanced by measuring and analyzing the interdisciplinarity of scientific research (Leydesdorff & Ivanova, 2021). Diversity, evenness and disparity are considered as three main dimensions of interdisciplinarity of research (L. Zhang et al., 2016). Specifically, diversity (also called variety) measures how many categories of disciplines there are in a study. In bibliometrics, the predefined disciplinary categories, such as the Leuven-Budapest subject classification system, the subject categories in Web of Science or Elsevier, has been used for calculating the diversity of research or journals, from the perspectives of citation or collaboration. Meanwhile, with the development of text mining techniques such as topic modeling, entity extraction and text clustering, the research topics, the specialty of authors has also been counted to represent the diversity of research (Bu et al., 2018). The interdisciplinary score of a research gets higher if the measurement object is more diverse in the categories. Although simple and clear, these indicators suffer from the redundant classification system, the synonyms of entities, and inaccurate topic clustering (Zuo & Zhao, 2018). From the perspective of networks, the between centrality has been calculated in co-citation networks and co-author networks to measure the diversity of research or journals (L. Zhang et al., 2016). However, these methods considered only the topological information of networks and ignored the nodal information. Besides, the diversity of research has also been examined from other perspectives, such as different regions, different h-index, different countries and different term functions (Bu, Ding, Xu, et al., 2018; Lu et al., 2019).

Evenness (also called balance or symmetry) measures how much of each category of discipline there is in a research. The evenness has a positive relationship with the diversity, that is, the more even the evenness, the higher the diversity. In fact, evenness and diversity are often treated as a "dual-concept" index, and the evenness of research was always calculated by the normalization of the diversity (Leydesdorff & Ivanova, 2021). For example, Corrêa Jr. et al. (2017) and Lu et al. (2019) employed the Shannon's entropy to normalize the diversity of author contributions and keyword term function for their calculating the evenness. Except for the Shannon's entropy, the Gini index and Simpson index have also been used for capturing the evenness in previous studies (L. Zhang et al., 2016). Finally, disparity quantifies how different from each other are the disciplinary categories in a research. It also has a positive relationship with the diversity. The Rao-Stirling diversity index was a typical measure for the disparity based on network structure. However, it was demonstrated to be not convincing for measuring the interdisciplinarity of journals by (Leydesdorff & Rafols, 2011) and (Zhou et al., 2012). The distance of disciplines that were operationalized by the cosine distance of topic terms or texts was also employed to quantify the disparity. For example, (Bromham et al., 2016) developed a co-classification-based indicator to measure the disparity. In the citation or collaboration networks, the shortest distances between nodes are also used for quantifying the disparity.

In this paper, at the macro level, we mainly discuss the diversity and the symmetry of research for the translation of drugs. Specifically, for a specific drug, we employ a Shannon entropy diversity (i.e., the true diversity) based on a bipartite graph that links the research categories to the drug's articles list, to quantify the diversity of research of the drug. Then the normalization of the diversity can be used to measure the symmetry of research of the drug, which indicates how even the drug research is. We choose this method because the similar bibliometric indicators have been successfully used for measuring the diversity and symmetry of the contributions of authors (Leydesdorff & Ivanova, 2021) and the term function of author keywords (Lu et al., 2019).

At the micro level, we also use two indicators (including the persistence of research and the stability of research) to analyze the patterns and the roles of interdisciplinary research in drug research. In the previous studies, Bu et al. (2018) proposed two bibliometric indicators, i.e., the Persistent Scientific Collaboration (based on the intervals and the skip years without collaborations) and the Stability of Collaboration (based on the year-to year publication output of collaborations), to measure the persistence and stability of collaborations for an author pair over time. The results of their studies on a large-scale dataset demonstrated the reliability and effectiveness of the two indicators (Bu, Ding, Liang, et al., 2018; Bu, Murray, Ding, et al., 2018). Therefore, in this paper, we replace the publication output of a specific author pair with the publication output of the specific category of research for a drug, and used the intervals and skip years without the specific category research (the Persistence of the Research) and the year-to-year publication output of the specific category of research (the Stability of Research) to quantify the persistence and stability of different categories of drug research, respectively.

## 2.2. Studies on translational research for drugs

Translational research for drugs (also known as translational pharmacology) is defined as the process of translating the drug discoveries in laboratory into clinical applications for human diseases (Kumar & Sattigeri, 2018). Despite the significant investment in drug research, research findings at molecular or animal levels are far from being fully translated to patient level (Contopoulos-Ioannidis et al., 2008; Madlock-Brown & Eichmann, 2015; Seyhan, 2019). Much of recent studies on translational pharmacology have been focused on identifying the key factors affecting this bench-to-bedside process (Rocca, 2017; Segura-Bedmar & Raez, 2019; Seyhan, 2019; Stubbs & Uzuner, 2019). For examples, Ioannidis (2016) systematically analyzed the reason for "useless clinical studies" and pointed out that a number of research findings published in medical journals are not robust as the authors has claimed and can't be reproduced mainly because of their poor study design. Fajardo-Ortiz et al. (2014) explored how the structure of knowledge of research teams affects the translation of a cancer drug (liposomal doxorubicin). Seyhan (2019) reviewed the challenges facing translational pharmacology and suggested that a number of "culprits" lead to "the lost in translation", such as inappropriate research hypothesis, irreproducibility of biomedical studies, misuse of statistical methods (i.e., p-value), and insufficient understanding of translational science. Agache et al. (2019) found that insufficient research funding and the reward mechanism were also important factors hindering the translation of drugs in the field of allergy. Dueñas et al. (2016) considered that most of the NIH funding often funds the small studies because they can be finished in a short time and get results published rapidly, while long-term studies with large patient cohorts sometimes can't be conduct because of insufficient resources available to the scientists. Almost all these studies recommended that interdisciplinary research, especially those involving both basic science and clinical science, should be encouraged to accelerate the "bench to bedside" process. These studies, however, were generally based on the qualitative analysis or manually systematic review. They have not yet to attempted to address the translational roadblocks by improving our understanding of the relationships between the interdisciplinarity of drug research and the translation of drugs.

**2.3. Studies on literature-based translational research in biomedicine**

With the advances in data availability and text analysis technologies, literature-based translational research in biomedicine has also become a popular subject in the field of bibliometrics and health informatics. Natural language processing has facilitated these bibliometric studies by enabling biomedical entity recognition (Xu et al., 2020), entity relationship reasoning (Lee et al., 2019) and entitymetrics (Li et al., 2020; Yu et al., 2021), etc. Researchers have also explored how to assess or track the translational progress of biomedical studies using automated literature mining. For example, Lewison and Paraje (2004) employed the clue words in the article titles to distinguish "basic" journals from "clinical" journals. Tijssen (2010) proposed a knowledge utilization triangle and classified the journals into six application domains according to their application orientation. However, these studies were not enough to represent the translational progress of articles. To go further, Boyack et al. (2014) trained a used a logistic regression model with article titles, abstracts and references to group more than 25 million articles from Scopus into different research levels. Meanwhile, MeSH terms indexed by experts have been widely used for tracking translational research. For example, Petersen et al. (2016) proposed a triple helix framework based on the "C", "D" and "E" groups of MeSH terms to describe the evolution of research focus over time. Weber (2013) developed a biomedicine triangle to identify translational science in three dimensions (i.e., "animal", "cell/molecular" and "human"). In addition, Hutchins et al. (2019) combined weber's biomedicine triangle and citation network information to predict the translational progress of biomedical studies from the perspective of knowledge flow. They pointed out that distinct knowledge pathways are significantly associated with the success of translation.

## 3. Methodology

To explore the interdisciplinarity of drug research from the perspective of translational science, we propose a four-step research framework: (1) data collecting and pre-processing; (2) characterizing interdisciplinarity of drug research; (3) translational analysis; and (4) pattern analysis. (Illustrated in Figure 1)

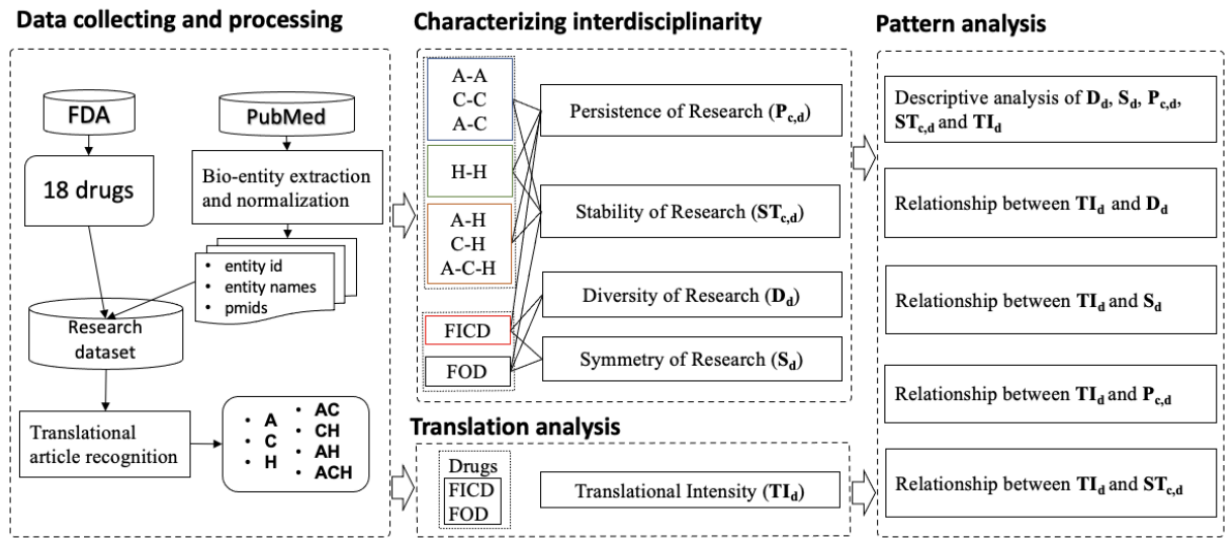

Figure 1. The overview of research design. (Note, A [Animal], C [cell/molecular], H [Human], FICD [first-in-class drug], FOD [follow-on drug].)

## 3.1. Data collecting and pre-processing

In this step, we first selected drugs for the analysis from the U.S. Food and Drug Administration (FDA) website (http://fda.gov/Drugs) based on two criteria: (a) drugs had to have been well-studied and approved by the FDA between the 1960's and 2000's; (b) each drug category had to have three (1 first-in-class and 2 follow-on) drugs included. Finally, 18 drugs belonging to 6 categories were selected: (1) angiotensin converting enzyme inhibitors (ACEIs), including captopril, enalapril and moexipril; (2) beta blockers (propranolol, atenolol and bisoprolol); (3) proton pump inhibitors (PPIs), containing omeprazole, lansoprazole and pantoprazole; (4) statins (lovastatin, simvastatin, and pitavastatin); (5) triptans (sumatriptan, almotriptan, and frovatriptan); and (6) nucleoside reverse transcriptase inhibitors (NRTIs), including zidovudine, lamivudine, and emtricitabine.

To obtain the related articles for these drugs, we downloaded over 30 million PubMed articles in XML format in August 2021. For each article, we extracted the bibliographic information such as title, abstract, MeSH terms and publication time using a dom4j-based XML parsing script. We next employed BioBERT (Lee et al., 2019), a biomedical language model pretrained on PubMed and PubMed Central, to extract biomedical entities (drugs) from the titles and abstracts. For each recognized drug entity, we assigned it a unique entity id with a multi-type biomedical entity normalization tool based on probability decision rules (Kim et al., 2019). 146,663 articles regarding these 18 drugs were finally filtered as research dataset and stored in a local MySQL database.

To track drugs along the translational continuum using aforementioned articles, it is crucial to distinguish basic articles from clinical articles. In this paper, we classified articles into Animal (A) related, Molecule/Cell (C) related, Human (H) related and combinations of these three (i.e., AC, AH, CH and ACH) using the MeSH terms assigned (Weber, 2013). Specifically, terms with MeSH codes starting with subcategory code A11, B02, B03, B04, G02.111.570 or G02.149 are C terms; terms with MeSH codes starting with subcategory code M01 or B01.050.150.900.649.801.400.112.400.400 are H terms; and terms with MeSH codes beginning with subcategory code B01 except for B01.050.150.900.649.801.400.112.400.400 are A terms. Each article could have one or more of the three kinds of MeSH terms, or none of the three (Hutchins et al., 2019; Weber, 2013). According to the article type, we divided research of drugs into 7 categories, i.e., AA, CC, and AC (research within basic science); HH (research within clinical science); CH, AH and ACH (interdisciplinary research involving both basic science and clinical science). The details on the research dataset are shown in Table 1.

In order to explicitly understand the interdisciplinarity in drug translational research, for each drug, we searched the PubMed (https://pubmed.ncbi.nlm.nih.gov/) and Derwent Innovation Index for the year of the earliest articles with different research types, the year of the first FDA approval, and the year of the first awarded patent. We also searched the Wikipedia and DrugBank (Wishart et al., 2018) to cross-verified this information, which are used to form the timelines of milestones for the 18 drugs. Finally, these information are mapped onto the Triangle of Biomedicine (Hutchins et al., 2019; Weber, 2013) for analysis.

Table 1. Overall information about the 18 drugs.

| Category | Drug Name | Level | # of articles in PubMed | AA | CC | HH | AC | CH | AH | ACH | FDA approval year |
|---|---|---|---|---|---|---|---|---|---|---|---|
| ACEIs | captopril | FICD | 13,624 | 3,504 | 247 | 6,373 | 1,653 | 725 | 473 | 291 | 1981 |
| | enalopril | FOD | 8,847 | 1,886 | 93 | 4,863 | 752 | 406 | 254 | 124 | 1984 |
| | moexipril | FOD | 127 | 21 | 9 | 61 | 4 | 6 | 1 | 3 | 1995 |
| Beta blockers | propranolol | FICD | 45,682 | 14,753 | 893 | 15,176 | 9,562 | 2,864 | 1,217 | 967 | 1964 |
| | atenolol | FOD | 8,427 | 1,318 | 356 | 4,467 | 778 | 554 | 198 | 149 | 1975 |
| | bisoprolol | FOD | 1,829 | 87 | 49 | 1,238 | 74 | 117 | 44 | 35 | 1986 |
| PPIs | omeprazole | FICD | 13,129 | 873 | 282 | 5,234 | 997 | 4,247 | 262 | 342 | 1988 |
| | lansoprazole | FOD | 3,042 | 91 | 116 | 1,185 | 153 | 1,142 | 48 | 95 | 1992 |
| | pantoprazole | FOD | 2,217 | 71 | 98 | 1018 | 83 | 527 | 39 | 64 | 1994 |
| Statins | lovastatin | FICD | 12,812 | 1,274 | 253 | 5,317 | 2,189 | 1,916 | 339 | 863 | 1987 |
| | simvastatin | FOD | 11,437 | 1009 | 208 | 5,412 | 1,289 | 1,411 | 328 | 593 | 1992 |
| | pitavastatin | FOD | 1,054 | 72 | 18 | 390 | 117 | 135 | 43 | 100 | 2003 |
| Triptans | sumatriptan | FICD | 3,363 | 351 | 32 | 2,047 | 216 | 139 | 165 | 103 | 1991 |
| | almotriptan | FOD | 301 | 8 | 6 | 241 | 2 | 5 | 18 | 4 | 2000 |
| | frovatriptan | FOD | 253 | 9 | 5 | 176 | 1 | 0 | 19 | 4 | 2001 |
| NRTIs | zidovudine | FICD | 12,673 | 246 | 456 | 4,356 | 631 | 5,793 | 162 | 571 | 1987 |
| | lamivudine | FOD | 10,513 | 67 | 227 | 3,318 | 157 | 5,613 | 83 | 268 | 1995 |
| | emtricitabine | FOD | 3,315 | 22 | 79 | 947 | 51 | 1,256 | 34 | 73 | 2006 |
| | *Total* | | *152,645* | *25,662* | *3,427* | *61,819* | *18,709* | *26,856* | *3,727* | *4,649* | |

### 3.2. Characterizing interdisciplinarity of drug research

In this section, we propose four indicators to quantitatively characterize interdisciplinarity of drug research from the perspective of translational science, that is, the diversity of research (**$D_d$**), the symmetry of research (**$S_d$**), the persistence of research (**$P_{c,d}$**), and the stability of research (**$ST_{c,d}$**). The $D_d$ and $S_d$ measure the diversity and evenness of the distribution of research of a given drug **d** at the macro level, respectively; while the **$P_{c,d}$** and **$ST_{c,d}$** quantify the persistence and stability of a kind of research **c** ( $c \in CT$, $CT =$

$\{AA, CC, HH, AC, AH, CH, ACH\}$) of a given drug **d** at the micro level. The detailed explanations for these indicators are as follows.

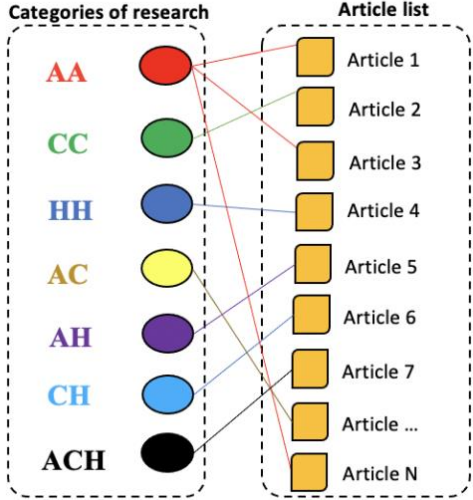

Figure 2. An example of a bipartite graph representing the relationship between drug related articles and the category of drug research. Note, the total number of articles assigned to particular research categories vary according to drugs.

**Diversity of research ($D_d$).** To quantify the diversity of research of a drug from the perspective of translational science, for each drug, the relationship among related articles and their research categories is treated as a bipartite graph (Corrêa Jr. et al., 2017; Lu et al., 2019), which is a graph with edges linked only among nodes belonging to two distinct groups. As illustrated in Figure 2, the bipartite graph derived from each drug establishes edges between drug related articles and their possible research categories. Each article is linked to one research category, while one research category can have multiple articles assigned.

For a specific research category $c$ ($c \in CT$, $CT = \{AA, CC, HH, AC, AH, CH, ACH\}$), its strength in the research of a drug, $I_c$, is given by:

$$I_c = \frac{\sum_i \omega_{ci} M_{ci}}{\sum_c \sum_i \omega_{ci} M_{ci}} \quad (1)$$

where $M_{ci}$ denotes the relationship between the research category $c$ and the i-th article in the research of a drug, i.e., if the research category of i-th article is $c$, $M_{ci} = 1$; else, $M_{ci} = 0$. Meanwhile, according to the research type of each article, we use the weight $\omega_{ci}$ to represent its importance to the drug research, i.e., if $c =$ *AA or CC or HH*, $\omega_{ci} = 1$; if $c =$ *AC or AH or CH*, $\omega_{ci} = 2$; else, $\omega_{ci} = 4$. Notably, the range of $I_c$ is from 0 to 1, and we can use the entropy concept to quantify the diversity of the distribution of the different categories of research for a drug (Corrêa Jr. et al., 2017; Lu et al., 2019). Thus, for a given drug $d$, the diversity of its research, $Diversity(D_d)$, is expressed as:

$$Diversity\ (D_d) = \exp\left(-\sum_{c \in CT} I_c \log I_c\right) \quad (2)$$

**Symmetry of research ($S_d$)** for a specific drug d means the evenness of the distribution of the strength for different categories of research of a drug. In this paper, we use the normalization of the diversity of research ($D_d$) to calculate how even the research of the drug (Corrêa Jr. et al., 2017; Lu et al., 2019). Therefore, for a given drug $d$, the symmetry of its research, $Symmetry\ (S_d)$, is calculated as:

$$Symmetry\ (S_d) = \frac{D_d}{n_{CT}} \quad (3)$$

where $n_{CT} (1 \leq n_{CT} \leq 7)$ denotes the number of the categories of research of a drug. $Symmetry\ (S_d)$ measures the symmetry of research of a drug and it ranges between 0 and 1. If different categories of research occur equally for a drug, $S_d$ will reach its maximum value, i.e., $S_d = 1$.

From the translational science perspective, we extend the two author-level bibliometric indicators (i.e., the degree of PSC and the stability of scientific collaboration) originally proposed by Bu et al. (Bu, Ding, Liang, et al., 2018; Bu, Murray, Ding, et al., 2018), to quantify the persistence and stability of the 7 kinds of drug research.

**Persistence of research ($P_{c,d}$)** is defined as the continuity of a specific category of research $c$ ($c \in CT$, $CT = \{AA, CC, HH, AC, AH, CH, ACH\}$) of a given drug $d$. We calculate $P_{c,d}$ by considering the number of years without $c$ research and whether these years are adjacent or not. Specifically, for a given drug $d$, we assume that the total number of $c$ research in $Y$ years is $P_c$, and we use a vector $\overrightarrow{P_c} = (p_{c,1}, p_{c,2}, \dots p_{c,q}, \dots, p_{c,y})$ to represent the yearly number of $c$ research among $Y$-year time, in which $\sum_{q=1}^{Y} p_{c,q} = P_c$. We then define $\overrightarrow{P_{c,m}}$ ($m \leq Y$) based on three criteria, i.e., $\overrightarrow{P_{c,m}}$ have to (1) be a consecutive sub-vector of $\overrightarrow{P_c}$; (2) contain $\mu$ components ($1 \leq \mu \leq Y$); (3) cater to the following restrictions ($1 < x_1 < x_2 \dots < x_\mu \leq Y$):

$$\begin{cases} p_{c,x_1} = p_{c,x_2} = \dots = p_{c,x_\mu} = 0 \\ p_{c,x_1-1} \neq 0\ (x_1 \neq 1)\ OR\ x_1 = 1 \\ p_{c,x_\mu+1} \neq 0\ (x_\mu \neq Y)\ OR\ x_\mu = Y \end{cases} \quad (4)$$

Then, the number of sub-vectors($\overrightarrow{P_{c,m}}$) that meet the criteria in $\overrightarrow{P_c}$ is denoted as $b_c$ (i.e., max(m) = $b_c$), which essentially means the count of intervals without $c$ research within the given $Y$ years. On the other hand, we define the number of years without $c$ research ($p_{c,q} = 0$) among the $Y$-year time as $SY_c$. Therefore, for a given drug $d$, the persistence of $c$ research, $P_{c,d}$, is calculated by:

$$Persistence\ (P_{c,d}) = Y - SY_c + \theta b_c \qquad (5)$$

where $\theta$ ranges in the interval (0,1) and is a fit parameter for this model. $P_{c,d} \in (1, Y]$ and each category of research of a drug can only have one value of $P_{c,d}$.

**Stability of research (ST$_{c,d}$)** of a specific category of research $c$ in the studies of a given drug $d$ among $Y$-year time is defined as:

$$Stability\ (ST_{c,d}) = 1 - \frac{\sum_{q=1}^{Y-1}|p_{c,q+1} - p_{c,q}|}{(Y-1) \times [\max(p_{c,q}) + 1]} \qquad (6)$$

where the value of $ST_{c,d}$ ranges in $[\frac{1}{\max(p_{c,q})}, 1]$. Specifically, if the yearly number of $c$ research in the studies of drug $d$ is a consistent number, then the value of $ST_{c,d}$ equals to 1; while, when the year number alternate between $\max(p_{c,q})$ for one year and zero the next year, the value of $ST_{c,d}$ will equal to $\frac{1}{\max(p_{c,q})}$.

### 3.3. Measuring the result of translation for a drug

The translation of a drug is a reiterative and continuous process (Seyhan, 2019). Many basic drug discoveries, such as captopril, aspirin and metformin (Spector et al., 2018), have been successfully translated to more than one human diseases. Hence, in this paper, we propose an indicator (i.e., translation intensity) to quantify the translation result of the whole research process of a drug, not limited to the first translation.

**Translation Intensity (TI$_d$)** measures the intensity to translate for the research of a drug. For a given drug $d$ with $N$ related articles in PubMed during the period of Y years, we employ the approximate potential to translate ($APT$), a machine-learning based indicator that measures the probability for an article to be cited in later clinical trials or guidelines (Hutchins et al., 2019), to quantify the translation intensity of each article related to the drug $d$. The APT of each article in PubMed can be freely obtained from the iCite website (https://icite.od.nih.gov/api) hosted by the NIH. Thus, the translation intensity of the drug $d$, $TI_d$, is expressed as:

$$TI_d = \frac{\sum_{a=1}^{N} APT_a}{Y} \qquad (7)$$

**3.4 Pattern analysis**

In this paper, we investigate the interdisciplinarity in drug research from the translational science perspective. Using 18 FDA approved drugs as research proxies, we first examine the first years of different kinds of research, the translation lag and the distribution of different kinds of research mapped onto the Triangle of Biomedicine from a bird's eye. Then, with the indicators proposed in this paper, we conduct descriptive analysis on the distribution patterns of the interdisciplinary characteristics of drug research, including the diversity of research, the symmetry of research, the persistence of research and the stability of research. We also describe the distribution of the translation intensity of the 18 drugs. The differences in these indicators are compared between all drugs, the first-in-class and the follow-on drugs as well as the 6 drug categories. Finally, we employ both the Pearson and Spearman correlation analysis to analyze the relationships between the translation intensity of drug and the interdisciplinary characteristics of drug research.

## 4. Results
### 4.1. overview

The median of the translation lags (Contopoulos-Ioannidis et al., 2008) for the 18 drugs from the earliest PubMed article or patent to their first FDA approval is 14 years (interquartile range, 7 to 17 years), with essentially no difference between the first-in-class drugs and the follow-on drugs. In fact, the actual translation lag could be longer because of technical protection and trade secrets by pharmaceutical companies. For example, captopril was originally synthesized in 1975, but until 1977, it was first awarded patent; and until 1978, the first article on captopril was recorded in the PubMed. This indicates the long length of time to translate basic discoveries into clinical applications of drugs, which confirms the findings of (Spector et al., 2018) and (Weber, 2013).

Figure 3 shows the first research of different categories, the first patents and the first FDA approvals along the timelines of the 18 drugs, which reveals three important findings. First, in drug research, intra-disciplinary research (e.g., AA and HH) occurred much earlier than interdisciplinary research involving both basic science and clinical science, such as AH, CH or ACH. Frequently, multidisciplinary research (ACH) appeared later than the first FDA approval of the drug, for examples, the beta blockers, the triptans and the NRTIs. Second, the sequences of first research in the timelines vary according to drugs. For instance, the first CC research of three triptans appeared at the last place of their timelines, while that of lamivudine ranked the first place. The differences among categories are larger than these of drugs in the same drug category. These differences in the structures of research among drugs may be caused by the intellectual structure and research progress of different drugs. Third, there is no significant difference in patterns between the first-in-class drugs and the follow-on drug.

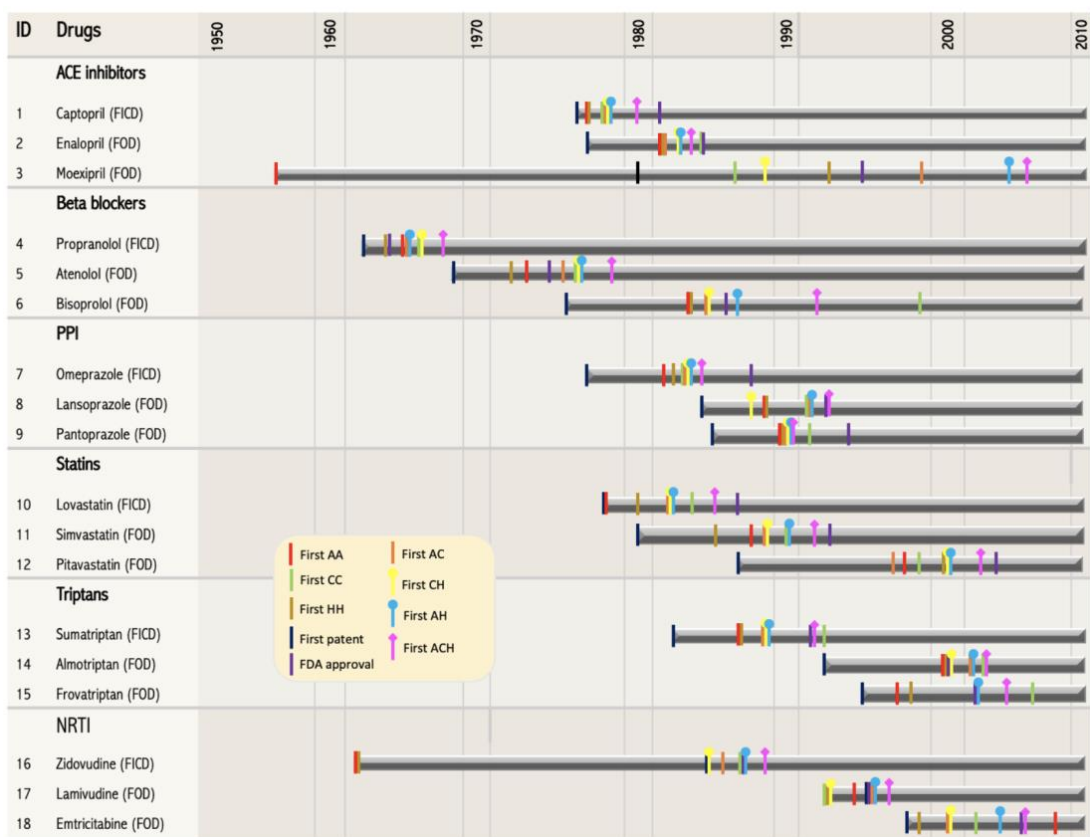

Figure 3. The first research of different categories, the first patents and the first FDA approvals mapped onto the timelines of the 18 drugs.

In Figure 4 and Figure 5, we mapped different categories of research of the 18 drugs during 1950-2018 from the perspective of translational science using the Triangle of Biomedicine originally proposed by Weber (2013) and modified by Hutchins et al. (2019), in which the three dimensions (Animal, Cell/molecular, and

Human) were scored into a Cartesian coordinate system and the research (articles) were represented by circles. The colors filled in circles vary according to the categories of research; the size of circles can reflect the number of studies (articles) sharing the same coordinate; and the location of a circle can indicate the A-C-H composition of the research of the drug. More details on the Triangle of Biomedicine can be found in (Hutchins et al., 2019; Weber, 2013).

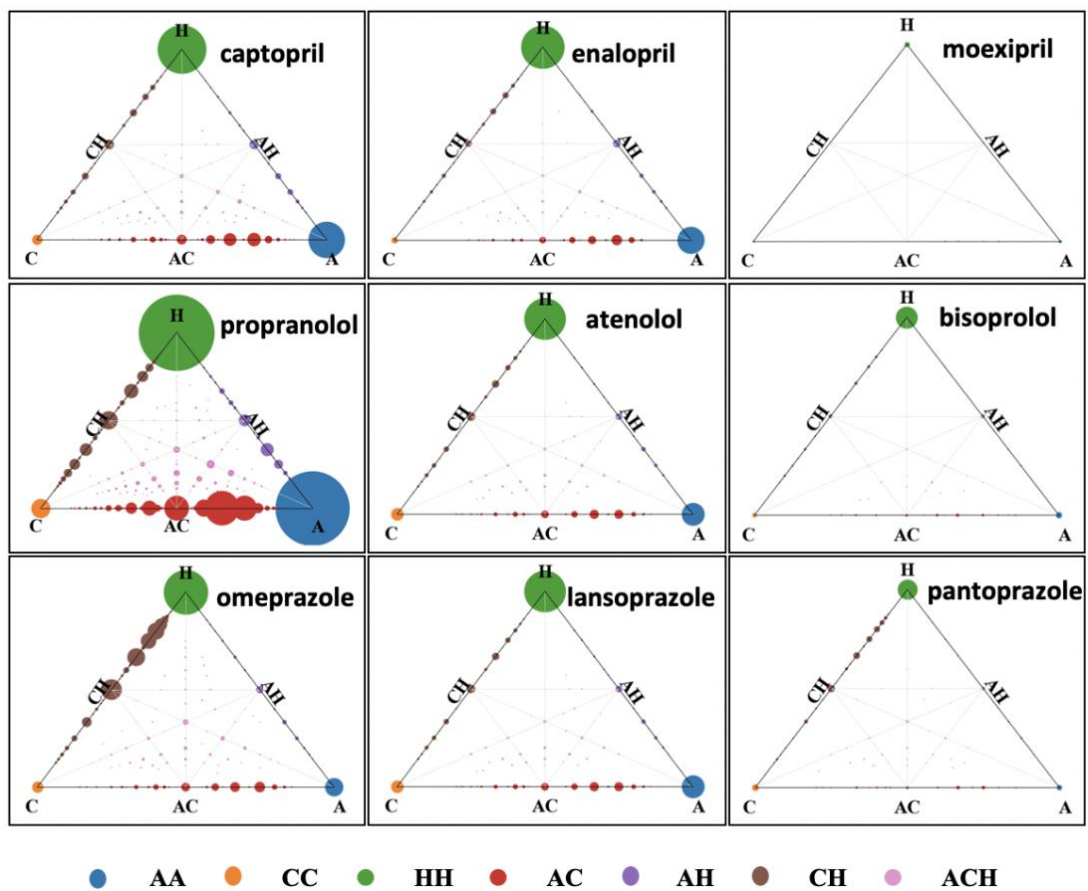

Figure 4. Different categories of research of ACE inhibitors, Beta blockers and PPIs mapped onto the Triangle of Biomedicine during the period of 1950-2018.

We note several general patterns when observing the distribution of research mapped onto the Triangle of Biomedicine in Figure 4 and Figure 5. First, for intra-disciplinary research, the total amount of HH research, whose mean percentage in all 18 drugs is 49.74% (interquartile range, 38.7% to 56.6%), possesses an absolute advantage over that of AA and CC research in all drugs. Nevertheless, knowledge discovered in basic science typically did not directly flows into human trials (Narin et al., 1976; Weber, 2013), which means discoveries in fundamental research was not successfully translated into clinical applications. Second, the total number of interdisciplinary research, especially for the research involving both basic science and clinical science (e.g., ACH), are much less than the intra-disciplinary research in drug research. Specifically, the total number of ACH research has been always ranking the last place for all drugs, and that of AH research is a clear second to the last. Besides, we also find that A and C are dominated in ACH research for all drugs when observing the positions of ACH research in the triangle.

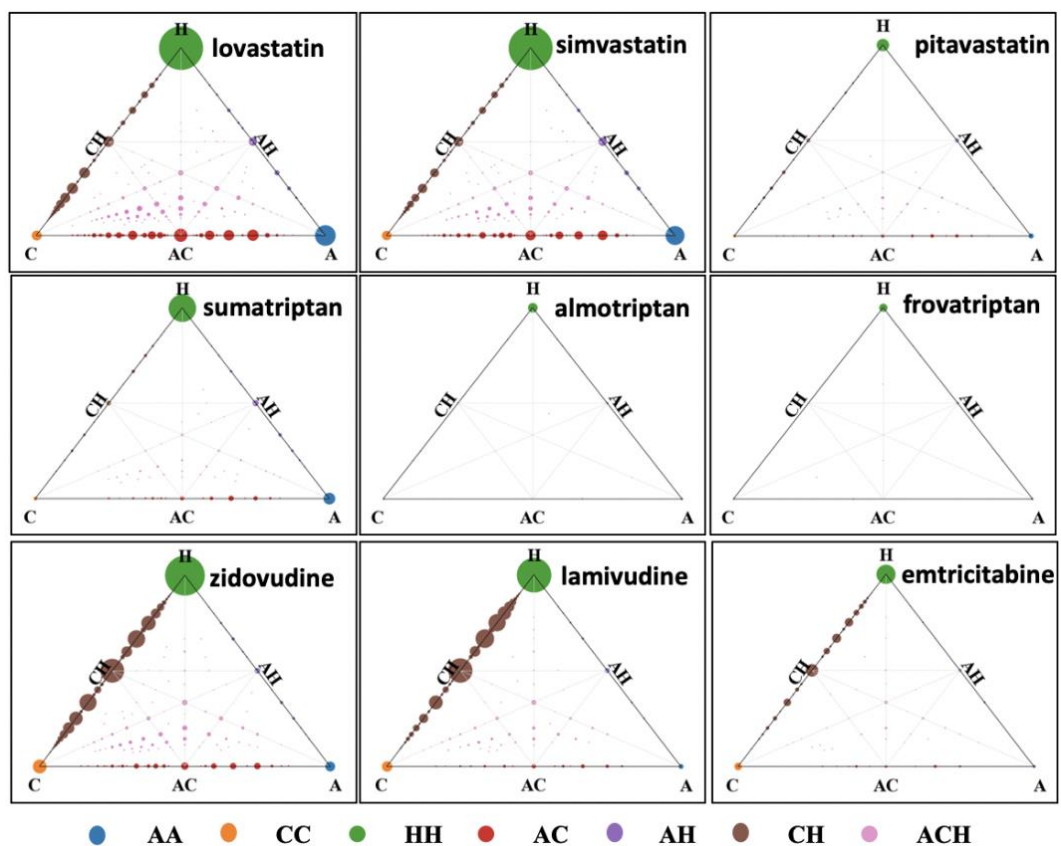

Figure 5. Different categories of research of Statins, Triptans and NRTIs mapped onto the Triangle of Biomedicine during the period of 1950-2018.

Third, the numbers of different categories of research of the first-in-class drugs are much larger than those of the follow-on drugs with the same drug category. This could be because the follow-on drugs "stand on the shoulders" of the first-in-class drugs and share progresses unveiled in the previous research of the first-in-class drugs. In addition, we can also find that the distributions of research for the first-in-class drugs are much more diverse.

The aforementioned observations raise questions about the relationships between the interdisciplinary characteristics of research and the translational research of drugs. For example, multidisciplinary research involving both basic science and clinical science could play a role in translational research for drugs through interdisciplinary ideas, resource sharing, science communication, talent training. Therefore, in the following sections, we further quantified the diversity, the symmetry, the persistence and the stability of research for drugs from the perspective of translational science, and we also examined the relationships between these four indicators with the translation intensity of the 18 drugs.

**4.2. Results of descriptive analysis**

*Diversity of research*

To explore how interdisciplinarity vary in drug research, we employed "diversity of research" ($D_d$) as a measure of the variability, as described in the methodology section. As shown in Figure 6, the red bars denote the diversity of research observed in the 18 drugs, with the grey bars as the reference when the category of research is ignored (i.e., $\omega_{ci}$ constantly equals to 1). We can make three observations. First, the diversity of research ($D_d$) varies according to drugs (mean, 4.33; and interquartile range, from 3.32 to 5.32). Specifically, the $D_d$ of propranolol (5.85) is highest among these 18 drugs, captopril (5.80) is a clear second, and moexipril (2.14) ranks at the last place. Second, the diversity of research ($D_d$) has a relative strong correlation with the drug category. The mean value of $D_d$ of Beta blockers (5.25) is much higher than Triptans (3.10). Third, the $D_d$ of the first-in-class drugs possess a significant advantage over the follow-on drugs across all six categories of drugs, confirming that the structures of research of the first-in-class drugs are much diverse than the follow-on drugs. Besides, all the deviations from the reference of the $D_d$ for all drugs are positive (the green arrows), indicating that interdisciplinary research contributes to the increase of diversity of drug research.

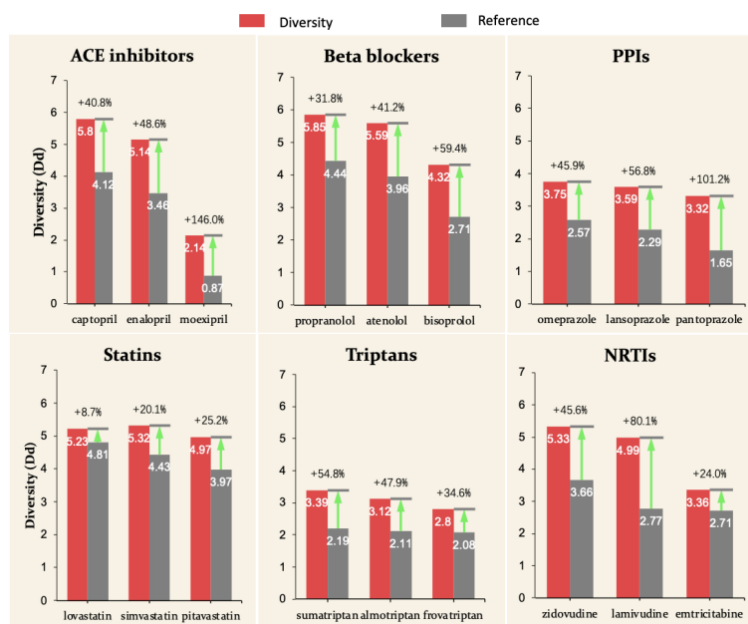

Figure 6. Diversity of research of the 18 drugs. Note, in each subgraph, the first drug is first-in-class drug and the following two are follow-on drugs. The grey bars are baselines when the category of research is ignored.

*Symmetry of research*

We examined the irregularity of research in drug research in terms of "symmetry of research" ($S_d$). As illustrated in Figure 7, we can find the corresponding value of $S_d$. The blue line represents the curve obtained by linking the values denoting the average symmetry of research of each drug, and the red one is the reference line when the category of research is ignored. Overall, the mean value of $S_d$ is 0.64 (interquartile range, from 0.45 to 0.76), indicating a lack of evenness in drug research, especially for moexipril (0.32) and frovatriptan (0.43). Meanwhile, the evenness of different categories of research in the first-in-class drugs is significantly better than the follow-on drugs in the same drug category by the value of $S_d$. However, PPIs is an exception, in which the $S_d$ of pantoprazole (0.57) is slightly larger than omeprazole (0.54). Moreover, the values of $S_d$ of all 18 drugs show a clear advantage over the references represented by the orange lines, although the values of $S_d$ vary according to drugs. Similarly, this demonstrates that interdisciplinary research can enhance the evenness of drug research.

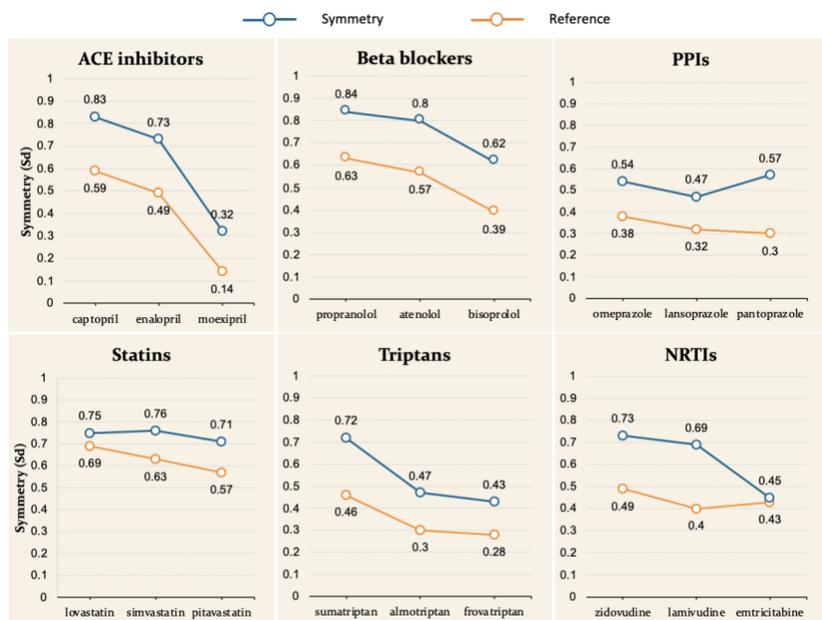

Figure 7. Symmetry of research of the 18 drugs. Note, in each subgraph, the first drug is first-in-class drug and the following two are follow-on drugs. The orange lines are baselines when the category of research is ignored.

*Persistence and stability of research*

In this section, we visualized the persistence and the stability of different categories of research for the 18 drugs using heatmaps. As shown in Figure 8a, the persistence of research of a given drug d is represented by a line comprised of seven cells, whose values (c, $P_{c,d}$) denote the value of the persistence ($P_{c,d}$) of the c research of the drug d. In our results, the degree of persistence for drugs is broad, ranging from 1.5 to 56. The colors of cells reflect the degree of persistence of research ($P_{c,d}$), that is, the higher the degree of $P_{c,d}$ is, the color is redder; or, the color will be bluer.

The visualization of persistence of drug research in Figure 8a led to several interesting findings. Overall, c the research of the first-in-class drugs, whose names are in bold, is evidently persistent than that of the follow-on drugs. In particular, for ACE inhibitors, the mean persistence of seven categories of research of the first-in-class drug captopril is 40.29, while those of the follow-on drugs are much lower, for example, 23.14 for enalopril and 10 for moexipril. As expected, the values of the persistence of research within disciplines (i.e., AA, CC and HH) are significantly higher than the research involving both basic science and clinical science (i.e., AH, CH and ACH). Furthermore, in research within disciplines, the persistence of AA or HH research has an advantage over that of CC research. This indicates that animal trials and human trials are important for the success of translation for drugs, although they are difficult to succeed and time-consuming. In addition, it can be found that, in most cases, the persistence of ACH research is the lowest compared to other six categories of research of the same drug.

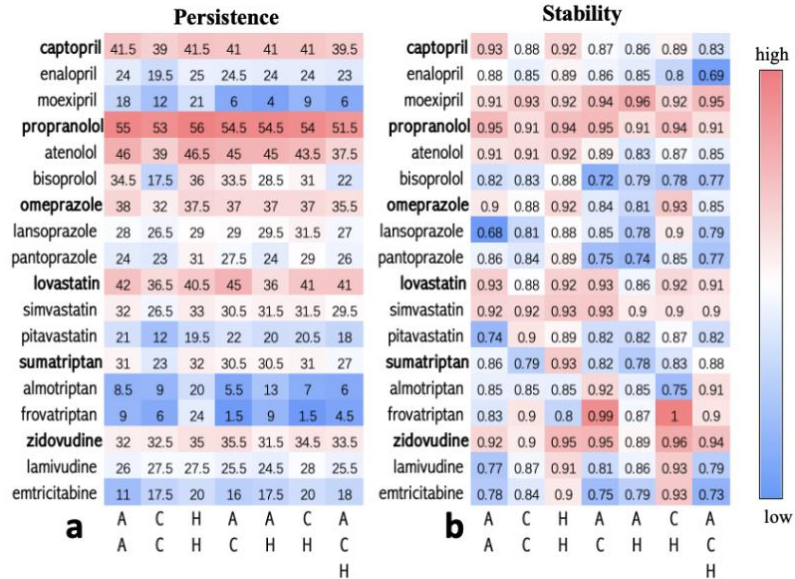

Figure 8. Persistence and stability of different categories of research for the 18 drugs. a. Heatmap of persistence of different categories of research for the 18 drugs. b. Heatmap of stability of different categories of research for the 18 drugs. Note, the name of the first-in-class drugs are in bold to distinguish from the follow-on drugs.

Similarly, as illustrated in Figure 8b, the stability of research of a given drug d is represented by a line comprised of seven cells, whose values (c, $ST_{c,d}$) denote the value of stability ($ST_{c,d}$) of the c research of the drug d. The heatmap shows that the values of $ST_{c,d}$ for the 18 drugs ranges from 0.68 to 1. The colors of cells represent the degree of the stability of research.

Overall, the values of $ST_{c,d}$ of research within disciplines (i.e., AA, CC and HH) possess an absolute advantage over the research involving both basic science and clinical science (i.e., AH, CH, and ACH). However, there are some exceptions, such as the stability of AA research of lansoprazole is the lowest (0.68), while that of CH research of frovatriptan is the highest (1). This could because that the basic issues on these two follow-on drugs have been well researched in the corresponding first-in-class drugs. Like the persistence of research, the stability of ACH research evidently ranks at the last place, and that of HH research shows the absolute advantage. However, there is no essential difference observed in the stability of research between the first-in-class drugs and the follow-on drugs.

***Translation Intensity***

We quantified the translation intensity ($TI_d$) by using the normalized sum of probabilities for all articles in the research of a drug d to be cited in later clinical trials or guidelines. The results are shown in Figure 9, in which the $TI_d$ of the first-in-class drugs is represented by the red bars for distinguishing from the follow-on drugs (in black). The scale of translation intensity of the 18 drugs (mean, 60.2; interquartile range, from 19.64 to 91.67) is also broad, ranging from 0.86 to 149.19.

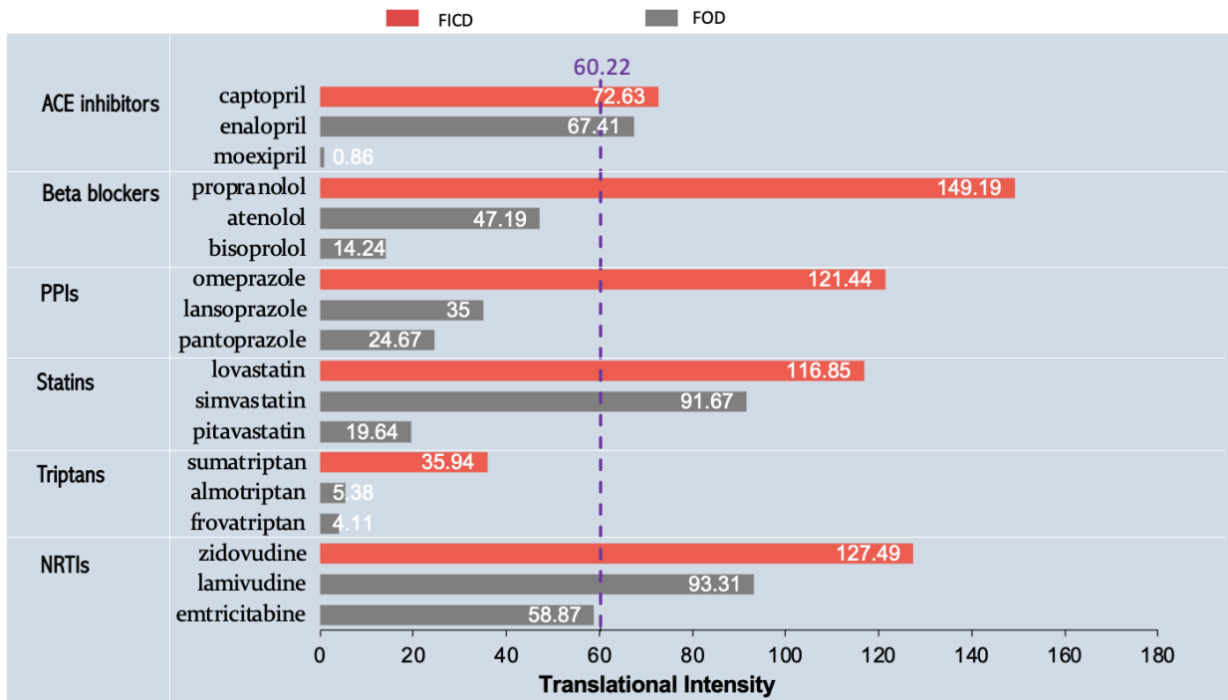

Figure 9. Translation intensity of the 18 drugs. Note, FICD [first-in-class drug] and FOD [follow-on drug].

Overall, the translation intensity of the 18 drugs varies according to drugs. Even for the drugs belonging to the same categories, there are significant differences between them. For example, the values of the translation intensity ($TI_d$) of ACE inhibitors are 72.63 for captopril, 67.41 for enalopril and only 0.86 for moexipril, respectively. Meanwhile, one can observe that, the first-in-class drugs (in red) usually have higher translation intensity values that surpassed the average level in most cases. This indicates that the research of the first-in-class drugs are more likely to be cited in clinical trials and guidelines, and they are easier translated into clinical applications.

The aforementioned findings we observed in the descriptive analysis drive us to ask a question: what are the relationships between the translation intensity and the interdisciplinary characteristics of (diversity, symmetry, persistence and stability) of drug research?

## 4.3. Relationships between the translation intensity and the interdisciplinary characteristics of drug research

In this section, the translation intensity of drugs was measured with the characteristics of interdisciplinarity in drug research. First, in terms of diversity of research ($D_d$), as presented in Figure 10a and Table 2, the result of correlation analysis shows a significant positive relationship between the diversity of research ($D_d$) and the translation intensity ($TI_d$): $r_{Pearson} = 0.719$, $p = 0.000$; $r_{Spearman} = 0.764$, $p = 0.000$.

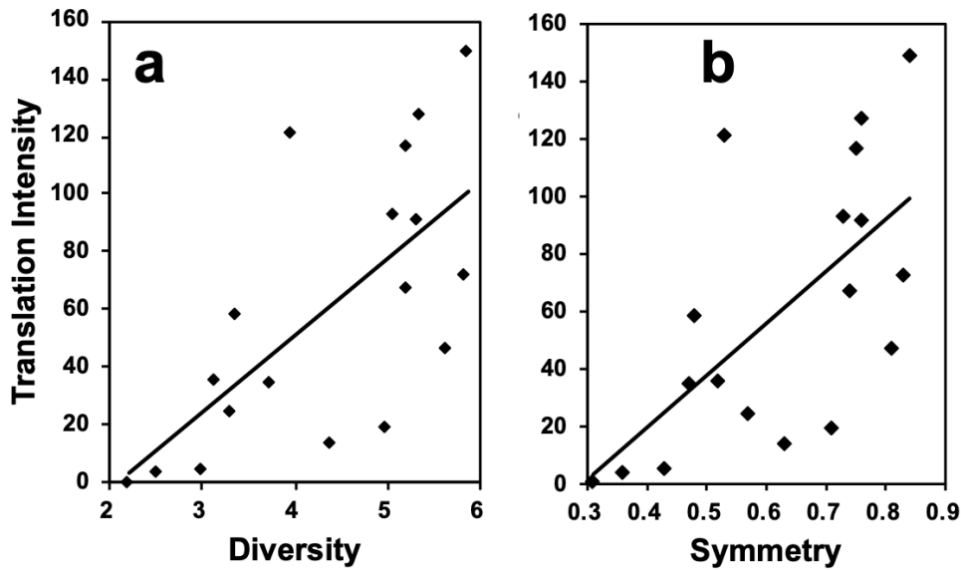

Figure 10. The relationship between the translation intensity and the diversity of research (a) or symmetry of research (b) in drug research.

Table 2. Correlation analysis of the translation intensity and the diversity (symmetry) of research in drug research.

|  | Pearson | p | Spearman | p |
|---|---|---|---|---|
| Diversity | 0.719 (***) | 0.000 | 0.777 (***) | 0.000 |
| Symmetry | 0.677(**) | 0.001 | 0.753 (**) | 0.001 |

(Note: $P < 0.05$, *; $P < 0.01$, **; $P < 0.001$, ***; two-tailed.)

Examining Figure 10b, we can observe that the value of translation intensity of drugs increases as the symmetry of research in drug research increases, indicating that there is also a relative strong positive association between the translation intensity ($TI_d$) and the symmetry of research ($S_d$) in drug research: $r_{Pearson} = 0.677$, $p = 0.001$; $r_{Spearman} = 0.753$, $p = 0.001$ (Table 2).

We then examined the translation intensity of drugs with the persistence ($P_{c,d}$) of seven different categories of drug research. As shown in Figure 11 and Table 3, there are significant positive relationships between the translation intensity ($TI_d$) and the $P_{c,d}$ of all seven kinds of drug research. As expected, the strength of the positive relationship between the translation intensity and the persistence of the ACH research is quite impressive: $r_{Pearson} = 0.807$, $p = 4.13*10^{-5}$; $r_{Spearman} = 0.792$, $p = 0.000$. Notably, the persistence of the CC research has a significant correlation with the translation intensity of drugs, with Pearson correlation coefficient high at 0.818 ($p = 0.000$) and Spearman correlation coefficient high at 0.835 ($p = 5.7*10^{-5}$). Moreover, the relationships between the translation intensity of drugs and the persistence of interdisciplinary research (AC, AH, CH and ACH) are stronger than those between the translation intensity of drugs and the persistence of research within disciplines (AA, CC and HH). In addition, the persistence of research involving both basic science and clinical science (AH, CH and ACH) has a stronger relationship with the translation intensity than the interdisciplinary research within basic science (AC). Eventually, when comparing the results in Table 2 and Table 3, we find that the persistence of interdisciplinary research (AC, AH, CH and ACH) has stronger correlation with the translation intensity than the diversity and the symmetry of research.

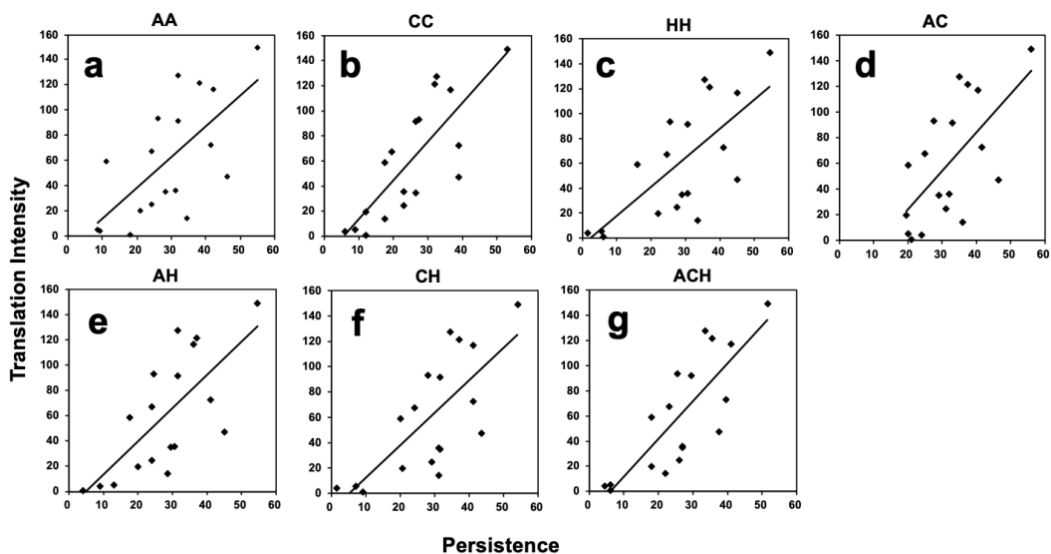

Figure 11. The relationship between the translation intensity of drugs and the persistence of drug research.

Table 3. Correlation analysis of the translation intensity of drugs and the persistence (Stability) of drug research.

|     | Persistence |  |  |  | Stability |  |  |  |
| --- | --- | --- | --- | --- | --- | --- | --- | --- |
|     | Pearson | $P$ | Spearman | $P$ | Pearson | $P$ | Spearman | $P$ |
| AA  | 0.663 (**) | 0.003 | 0.691 (**) | 0.000 | 0.465 | 0.316 | 0.543 | 0.071 |
| CC  | 0.818 (***) | 0.000 | 0.835 (***) | $5.7*10^{-5}$ | 0.246 | 0.137 | 0.193 | 0.053 |
| HH  | 0.637 (**) | 0.002 | 0.666 (**) | 0.002 | 0.497 | 0.057 | 0.377 | 0.130 |
| AC  | 0.715 (**) | 0.001 | 0.729 (**) | 0.001 | 0.241 | 0.400 | 0.160 | 0.029 |
| AH  | 0.738 (**) | 0.001 | 0.752 (***) | 0.000 | 0273 | 0.278 | 0.242 | 0.598 |
| CH  | 0.728 (**) | 0.001 | 0.714 (**) | 0.001 | 0.437 | 0.196 | 0.433 | 0.812 |
| ACH | 0.807 (***) | $4.13*10^{-5}$ | 0.792 (***) | 0.000 | 0.163 | 0.821 | 0.097 | 0.222 |

(Note: P < 0.05, *; P < 0.01, **; P < 0.001, ***; two-tailed.)

The relationships between the translation intensity of drugs and the stability ($ST_{c,d}$) of seven different kinds of drug research were finally explored, noted in Figure 12 and Table 3. We can see that, although there are trends for positive relationships between the translation intensity and the stability of research, these relationships are not statistically significant.

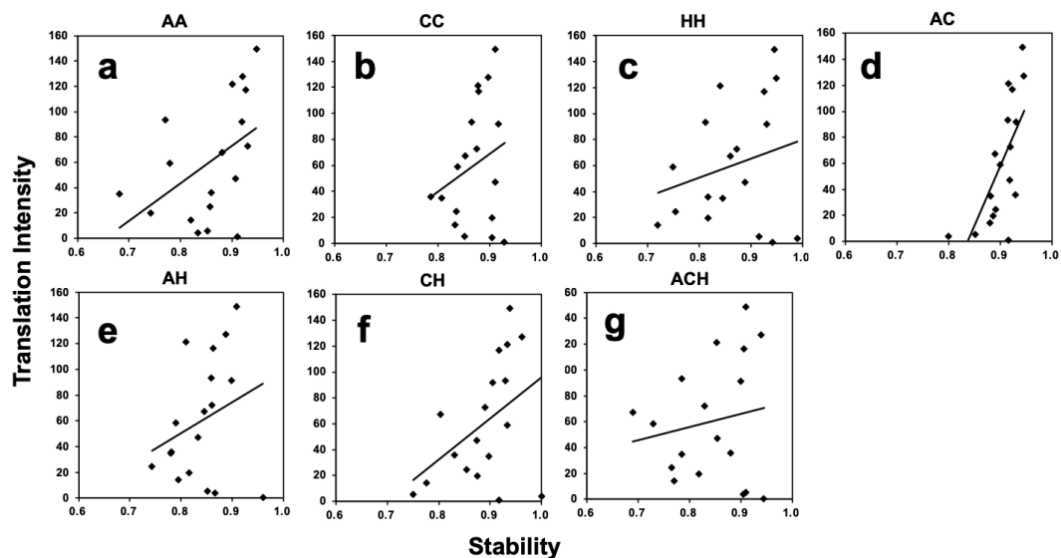

Figure 12. The relationship between the translation intensity of drugs and the stability of drug research.

## 5. Discussion and conclusion

Translational pharmacology is today playing an important role in drug development by linking basic discoveries to clinical needs (Dolgos et al., 2016; Kumar & Sattigeri, 2018). In this paper, we analyze the interdisciplinarity in drug research from a novel bibliometric perspective, i.e., the translational science perspective. 18 FDA approved drugs belonging to 6 categories are used as the research proxies. Specifically, we propose four bibliometric indicators (i.e., the Diversity of Research, the Symmetry of Research, the Persistence of Research and the Stability of Research) for characterizing interdisciplinarity in drug research at both the macro and the micro levels. We also design a bibliometric indicator (i.e., Translation Intensity) for quantifying the degree of clinical translation for drugs. We analyze the relationships between the interdisciplinary characteristics of drug research and the translation intensity of drugs.

We find that it takes a long length of time to translate a laboratory discovery for human disease (Figure 3). Interdisciplinary research involving both basic science and clinical science have been suggested in many previous studies to accelerate this process ( Banner et al., 2019; Kumar & Sattigeri, 2018; Seyhan, 2019); however, they (especially ACH research) are still rare in the research of all the 18 drugs (Figure 4 and 5) and they generally occurred later than intra-disciplinary research in the timelines (Figure 3). Furthermore, the descriptive analysis indicates that the interdisciplinary characteristics of drug research and the translation intensity of drugs both vary and show different patterns according to drugs (Figure 6, 7, 8, and 9), raising the question about the relationships between these interdisciplinary characteristics of drug research and the translation intensity of drugs. Correlation analyses show that several interdisciplinary characteristics of drug research, such as the diversity of research and the persistence of ACH research, have significantly influenced the translation of drugs.

Actually, several research in scientometrics have demonstrated that the importance of the diversity to radical innovations, academic impact, and the success of scientific career (Amjad et al., 2017; Gil-Garcia et al., 2019; Xu et al., 2015). High diverse research and collaborations can promote translational research because of more financial support (Seyhan, 2019) as well as talents who come from various cultures, backgrounds and experiences (Xu et al., 2015). We find that drugs whose research have higher diversity and symmetry are usually those more successful drugs (i.e., the first-in-class drugs) with higher values of translational intensity, such as propranolol and captopril, indicating that the diversity and symmetry of research have contributed to enhance the translation of basic discoveries in drug research.

Another factor that has influenced the translation intensity of drug is the persistence of research. The persistence of all kinds of research has positive correlations with the translation intensity of drugs, and the relationship between the CC research and the translation intensity is very impressive. These findings verify that the old saying "success lies in perseverance" also applies to the field of drug research. The persistent efforts in basic research at cell/molecular level is significant for the translation of drugs. Meanwhile, the persistence of interdisciplinary research, especially those involving both basic and clinical science, have stronger relationships with the translation intensity of drugs than those of intra-disciplinary research. This can be explained by the finding of (Bu, Ding, Liang, et al., 2018; Bu, Ding, Xu, et al., 2018) that interdisciplinary research needs a high degree of persistence to generate high-impact outputs, while the intra-disciplinary research requires only a moderate degree of persistence.

We find that the stability of research did not show a significant influence on the translation intensity of drugs. Compared with the persistence of research, the stability of research puts more emphasis on the changes of the number of research over time in drug research (Bu, Murray, Ding, et al., 2018; Ioannidis et al., 2014). This finding indicates that the persistence of research means a lot to the translation intensity of drugs, while the amount of research and its changes over time do not.

This paper has several implications. Methodologically, this study showcases how to investigate the interdisciplinarity in drug research from the perspective of translational science, which provides a novel bibliometric perspective for the research community and could be applied to other domains. Meanwhile, there are significant difference detected between breakthroughs (the first-in-class drugs) and the follow-on drugs by the 5 indicators proposed in this paper, including the 4 bibliometric indicators for characterizing interdisciplinarity of drug research and the translation intensity of drugs. This demonstrates that these 5 indicators could be valuable for pharmaceutical companies, policymakers and researchers to predict the success of drugs. In addition, the indicators proposed in this paper could be adopted and duplicated to other objects, such as medical devices, vaccines and human genes.

There are several limitations of this study. First, our analysis considers only seven kinds of research in drug research from the translation perspective, which ignores the research at other levels, such as research involving the industry, the academia, the funding agencies and the government departments. The landscape of research among these different agencies could have influenced the translation of drugs. Second, the data used in our analysis is limited to the PubMed articles. Some other data source on drug research and development, such as patents, clinical trials, government files and web pages, in which interdisciplinary information were recorded, should be included. In our future work, we will integrate different categories of data sources for investigating multiple attributes of drug research and take into consideration other types of research at other levels (e.g., author, institution and biomedical entities). In addition, these findings in this paper mainly based on the 18 FDA approved drugs; although these drugs belong to six different drug categories used for different diseases, we don't know whether they can be generalized to all drugs. In future work, we will aim to test the proposed measures on other drugs to see whether generalized patterns exist in different translation processes. we also plan to use data driven methods and regression analysis on a large-scale drug-related dataset to understand the role of different kinds of research and their interdisciplinary characteristics in translating basic drug discoveries to therapies for human diseases.

**Acknowledgement**

The support provided by the Major Projects of the National Social and Social Science Fund (No. 17ZDA292) and the China Scholarship Council (CSC) during the visit by Xin Li to the University of Texas at Austin (No. 201806270047) are acknowledged. The support provided by Wuhan University (student exchange program) during the visit by Xuli Tang to the University of Texas at Austin is acknowledged. The authors are grateful to the anonymous referees and editors for their invaluable and insightful comments. We would like to express special gratitude to Prof. Ying Ding for her valuable comments and editorial assistance.